\documentclass[aps,prl,amsmath,twocolumn,superscriptaddress,showpacs]{revtex4}

\usepackage{graphicx}
\usepackage{bm} 
\usepackage{url}

\DeclareMathOperator{\Img}{\mathrm{Im}}
\DeclareMathOperator{\Rea}{\mathrm{Re}}
\DeclareMathOperator{\Sp}{\mathrm{Sp}}
\begin{document}
\author{Mikhail S. Kalenkov}
\affiliation{I.E. Tamm Department of Theoretical Physics, P.N. Lebedev Physical Institute, 119991 Moscow, Russia}
\affiliation{Laboratory of Cryogenic Nanoelectronics, Nizhny Novgorod State Technical University, 603950 Nizhny Novgorod, Russia}
\author{Andrei D. Zaikin}
\affiliation{Institut f\"ur Nanotechnologie, Karlsruher Institut f\"ur Technologie
(KIT), 76021 Karlsruhe, Germany}
\affiliation{I.E. Tamm Department of Theoretical Physics, P.N. Lebedev Physical Institute, 119991 Moscow, Russia}
\affiliation{Laboratory of Cryogenic Nanoelectronics, Nizhny Novgorod State Technical University, 603950 Nizhny Novgorod, Russia}
\author{Leonid S. Kuzmin}
\affiliation{Chalmers University of Technology, Goetenburg, Sweden}
\affiliation{Laboratory of Cryogenic Nanoelectronics, Nizhny Novgorod State Technical University, 603950 Nizhny Novgorod, Russia}

\title{Theory of a large thermoeffect in superconductors doped with magnetic impurities}

\begin{abstract}
We argue that parametrically strong enhancement of a thermoelectric current can be observed in conventional superconductors
doped by magnetic impurities. This effect is caused by violation of the symmetry between electron-like and hole-like
excitations due to formation of subgap Andreev bound states in the vicinity of magnetic impurities. We develop
a quantitative theory of this effect and demonstrate that it can be detected in modern experiments.

\end{abstract}

\pacs{74.25.fg, 74.25.F-, 75.20.Hr}
\maketitle

Application of an electric field $\bm{E}$ to a normal
conductor with Drude conductivity $\sigma_N$ yields an electric current $\bm{j}=\sigma_N \bm{E}$ across this conductor. A similar effect
can be produced by a temperature gradient  $\nabla T$. In this case the current $\bm{j}$ induced in a sample takes the form $\bm{j}=\alpha_N \nabla T$, where
$\alpha_N \sim (\sigma_N/e)(T/\epsilon_F)$ is thermoelectric coefficient and $\epsilon_F$ is the Fermi energy. 
The latter simple equations illustrate the essence of the so-called thermoelectric effect in normal metals.

\begin{figure}
\centerline{\includegraphics[width=50mm]{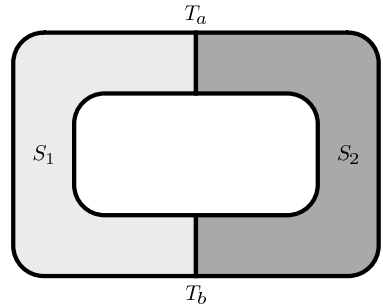}}
\caption{A ring formed by two different superconductors with contacts
maintained at different temperatures $T_a$ and $T_b$.}
\label{ss-fig}
\end{figure}

If a metal becomes superconducting, the situation changes significantly. On one hand, the electric field cannot anymore penetrate into 
a superconductor and, hence, the Drude contribution to the current is absent in this case. On the other hand, a supercurrent 
$\bm{j}_s$ can now be induced in the sample without any electric field. It follows immediately that by applying a
temperature gradient to a uniform superconductor one would not be able to induce
any current since thermal current would be exactly compensated by the
supercurrent $\bm{j}_s =- \alpha \nabla T$, where $\alpha$ defines thermoelectric
coefficient in a superconducting state. Ginzburg
\cite{Ginzburg44,Ginzburg91} demonstrated that no such compensation generally
occurs in non-uniform superconductors which opens a possibility to
experimentally detect thermoelectric current in such structures.
Several experiments with bimetallic superconducting rings (see Fig. \ref{ss-fig})
have been performed
\cite{Zavaritskii74,Falco76,Harlingen80} which indeed revealed the presence of
thermoelectric magnetic flux in such rings. However, both the magnitude of the
effect and its temperature dependence turned out be in a strong disagreement
with available theoretical predictions \cite{Galperin73}. Quite surprisingly, the
magnitude of the thermoeffect detected in these experiments exceeded theoretical
estimates by {\it several orders of magnitude}. Subsequently, a good agreement
between theory and experiment \cite{Gerasimov97} was claimed, but this report
remained largely unnoticed. In any case, no convincing explanation of the
discrepancy between experiments \cite{Zavaritskii74,Falco76,Harlingen80}
and theory \cite{Galperin73} was offered and the paradox remains unresolved
until now \cite{NL}.

In this Letter we are not aiming at directly resolving this long standing paradox. Rather our
primary goal is to identify the conditions under which
thermoelectric currents in superconductors can be significantly enhanced.
In the normal state contributions to the thermoelectric
coefficient $\alpha_N$  from electron-like and hole-like excitations are of
the opposite sign and almost cancel each other. A similar situation occurs in
conventional superconductors where the thermoelectric
coefficient $\alpha$ also remains small \cite{Galperin73} and monotonously decreases with $T$
below the critical temperature $T_c$.
On the other hand, in
unconventional superconductors impurity scattering may lead to much larger
values of $\alpha$ due to formation of quasi-bound Andreev states near
impurities which yield high asymmetry between electron and hole scattering
rates \cite{Arfi88,Lofwander04}.

Here we will demonstrate that ``giant'' thermoeffect can also be expected in
conventional superconductors doped by magnetic impurities. Also in this case
Andreev bound states are formed near such impurities
\cite{Shiba68,Rusinov68,Rusinov69} thereby explicitly breaking the symmetry
between electron and holes \cite{Kon80,Zaitsev86}. We argue that
this feature may cause parametrically strong enhancement of the thermoeffect in such systems
\begin{equation}
\alpha /\alpha_N(T_c) \sim p_F \ell \gg 1,
\label{est}
\end{equation}
where $p_F=mv_F$ is the Fermi momentum and $\ell$ is the electron elastic mean
free path in the absence of magnetic impurities. This formula remains valid in
the most relevant diffusive limit $\ell \lesssim v_F/T_c$ and at an ``optimal''
concentration of magnetic impurities $n_{\rm imp}$ roughly equal to one half of the critical one (see below).
Eq. (\ref{est}) predicts possible enhancement of the thermoeffect in superconductors with magnetic
impurities by several orders of magnitude as compared to that in the normal
state at $T=T_c$.

{\it Quasiclassical formalism and impurity self-averaging}.
In what follows we will consider a superconductor which contains both
non-magnetic and magnetic impurities. Our analysis is based on the
quasiclassical formalism of nonequilibrium Green-Keldysh matrix functions
$\check g$ obeying the Eilenberger equations \cite{bel}
\begin{equation}
-i\bm{v}_F\nabla \check g(\bm{p}_F,\bm{r},\varepsilon,t) =
[\check \Omega - \check \Sigma, \check g(\bm{p}_F,\bm{r},\varepsilon,t)],
\
\check g^2 =1.
\label{eilen}
\end{equation}
The check symbol denotes $4\times 4$ Keldysh matrices
\begin{equation}
\check X =
\begin{pmatrix}
\hat X^R & \hat X^K \\
0 & \hat X^A\\
\end{pmatrix},
\quad
X=g,\,\Omega ,\,\Sigma,
\end{equation}
with blocks $\hat X^{R,A,K}$ being $2\times2$ matrices in the Nambu space.
The matrix $\check \Omega$ has the standard structure
\begin{equation}
\hat \Omega^R = \hat \Omega^A =
\begin{pmatrix}
\varepsilon & \Delta \\
-\Delta  & -\varepsilon\\
\end{pmatrix},
\quad
\hat \Omega^K=0,
\end{equation}
where $\varepsilon$ is the quasiparticle energy, $\Delta$ is the BCS order
parameter which is chosen real further below.

Scattering of electrons on impurities is
accounted for by the self-energy matrix $\check \Sigma$ which can be expressed
in the form
\begin{equation}
\check \Sigma = - i\Gamma \left<\check g\right> + \check \Sigma_{\rm m},
\quad
\Gamma=v_F/(2\ell).
\label{selfe}
\end{equation}
Here the first term describes the effect of non-magnetic isotropic
impurities while the second term  $\check \Sigma_{\rm m}$
is responsible for electron scattering on randomly distributed magnetic
impurities \cite{Rusinov69}
\begin{multline}
\check \Sigma_{\rm m}=\dfrac{n_{\rm imp}}{2\pi N_0}
\Bigl\{
\left( [u_1 + \hat \tau_3 u_2]^{-1} +
i \left< \check g \right> \right)^{-1}
+\\+
\left( [u_1 - \hat \tau_3 u_2]^{-1} + i \left< \check g \right> \right)^{-1}
\Bigr\},
\label{selfem}
\end{multline}
where $N_0$ is the electron density of states per spin direction at the
Fermi level, $u_{1,2}$ are
dimensionless parameters characterizing the impurity scattering potential and
$\hat \tau_3$ is Pauli matrix in the Nambu space.
Averaging over the Fermi surface is denoted by angular brackets
$\left<\cdots\right>$. Note that within the Born approximation the self-energy \eqref{selfem} 
just reduces to the well known Abrikosov-Gor'kov result
\cite{Abrikosov60}. Unfortunately this approximation is insufficient for our present purposes since it does not
allow to account for impurity Andreev bound states (impurity bands) and the electron-hole asymmetry. For this reason in what follows
we will go beyond Born approximation and employ a more general expression for the self-energy \eqref{selfem}.

Finally, the current density $\bm{j}$ is defined
with the aid of the standard relation
\begin{equation}
\bm{j}(\bm{r}, t)= -\dfrac{e N_0}{4} \int d \varepsilon
\left< \bm{v}_F \mathrm{Sp} [\hat \tau_3 \hat g^K(\bm{p}_F,
\bm{r},\varepsilon, t)] \right>.
\label{current}
\end{equation}

{\it Electron-hole asymmetry and the density of states.}
It is well known that two subgap Andreev bound states with
energies
\begin{equation}
\varepsilon_B=\pm\beta\Delta ,\quad \beta^2 = \dfrac{( 1 + u_1^2-u_2^2 )^2}{( 1 + u_1^2-u_2^2 )^2 + 4 u_2^2}.
\label{betaa}
\end{equation}
are localized near each magnetic impurity in a superconductor \cite{Shiba68,Rusinov68}.
Similarly to the case of unconventional superconductors
with non-magnetic impurities \cite{Arfi88,Lofwander04} these Andreev bound states 
yield different scattering rates for electrons and holes and, hence, break the
electron-hole symmetry in our system thereby causing strong enhancement of the thermoeffect.

\begin{figure}
\centerline{\includegraphics[width=80mm]{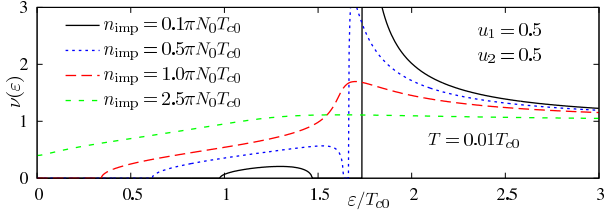}}
\caption{(Color online) Energy resolved density of states $\nu(\varepsilon)=\nu(-\varepsilon)$
in a superconductor doped by magnetic impurities. $T_{c0}$ is the critical temperature of an undoped superconductor.}
\label{dos-fig}
\end{figure}

Consider the retarded part of the self-energy $\check \Sigma$ \eqref{selfe}.
It can be written in the form
\begin{equation}
\hat \Sigma^R =
\begin{pmatrix}
\Sigma_0^R + \Sigma_g^R & \Sigma_F^R \\
\Sigma_{F^+}^R & \Sigma_0^R - \Sigma_g^R \\
\end{pmatrix},
\label{selferet}
\end{equation}
where non-vanishing diagonal part $\Sigma_0^R$ explicitly accounts for asymmetry between electrons and holes \cite{Zaitsev86}.
Substituting the retarded Green function matrix
\begin{equation}
\hat g^R=\dfrac{1}{ \sqrt{\strut \bar \varepsilon^2 - \bar \Delta^2}}
\begin{pmatrix}
\bar \varepsilon & \bar \Delta \\
-\bar \Delta & -\bar \varepsilon\\
\end{pmatrix}
\label{gbar}
\end{equation}
into Eqs. \eqref{selfe}, \eqref{selfem} we evaluate $\hat \Sigma_0^R$ as well as the energy resolved superconducting density of
states $\nu(\varepsilon)$ normalized to its normal state value. Introducing the parameter $\tilde \varepsilon = \bar
\varepsilon \Delta / \bar \Delta$ we get 
\begin{gather}
\Sigma_0^R(\varepsilon)=
\Gamma_0
\dfrac{\tilde \varepsilon^2 - \strut \Delta^2 }{
\tilde \varepsilon^2 - \beta^2 \strut \Delta^2 }, \quad \nu(\varepsilon)=\Rea\dfrac{\tilde \varepsilon}{\sqrt{\tilde \varepsilon^2 - \Delta^2}},
\label{nusigma0}
\\
\sqrt{\strut \bar \varepsilon^2 - \bar \Delta^2}=
\sqrt{\tilde \varepsilon^2 - \Delta^2} + i\Gamma +
i \Gamma_1
\dfrac{\tilde \varepsilon^2 - \Delta^2}{
\tilde \varepsilon^2 - \beta^2 \Delta^2 },
\label{sqrt}
\end{gather}
where the parameter $\tilde \varepsilon$ is fixed
by the relation \cite{Shiba68,Rusinov68}
\begin{equation}
\tilde \varepsilon = \varepsilon
+i\Gamma_2
\dfrac{\tilde \varepsilon \sqrt{\tilde \varepsilon^2 - \Delta^2}
}{\tilde \varepsilon^2 - \beta^2 \Delta^2},
\label{etilde}
\end{equation}
The scattering parameters $\Gamma_{0,1,2}$ have the dimension of rates being
proportional to the concentration of magnetic impurities $n_{\rm imp}$. They read
\begin{gather}
\Gamma_0=\dfrac{n_{\rm imp}}{\pi N_0}\dfrac{u_1 ( 1 + u_1^2-u_2^2 )}{( 1 + u_1^2-u_2^2 )^2 + 4 u_2^2  },
\label{Gamma0}
\\
\Gamma_1=
\dfrac{n_{\rm imp}}{\pi N_0}
\dfrac{( 1 + u_1^2-u_2^2 ) (u_1^2-u_2^2)}{( 1 + u_1^2-u_2^2 )^2 + 4 u_2^2  }
\label{Gamma1}
\\
\Gamma_2 =2\dfrac{n_{\rm imp}}{\pi N_0}\dfrac{u_2^2}{( 1 + u_1^2-u_2^2 )^2 + 4 u_2^2  }.
\label{Gamma2}
\end{gather}
Note that the parameters $\tilde \varepsilon$, $\Sigma_0^R$ and $\nu(\varepsilon)$ remain insensitive to the
electron scattering rate on non-magnetic impurities $\Gamma$ since such scattering does not produce any pair-breaking effect
in bulk conventional superconductors. On the contrary, scattering on magnetic impurities may strongly modify these parameters.
For illustration, the density of states $\nu(\varepsilon)$ is depicted in Fig. \ref{dos-fig} at $\varepsilon>0$
and different values $n_{\rm imp}$. With increasing $n_{\rm imp}$ Andreev levels get broadened forming two impurity bands
respectively at positive and negative energies. Further increase of $n_{\rm imp}$ yields even broader bands which eventually
merge with continuum (overgap) states.

{\it Thermoeffect enhancement by magnetic impurities.}
We are now prepared to evaluate the thermoelectric coefficient $\alpha$.
In doing so we will essentially follow the quasiclassical
linear response theory initially formulated in Ref. \onlinecite{Graf96}
for the analysis of thermal conductivity in unconventional superconductors. This approach allows to recover
the dominating contribution to the thermoelectric coefficient $\alpha$ which originates from the electron-hole
asymmetry. Employing Eqs. (\ref{eilen}) and proceeding along the lines with Ref. \onlinecite{Graf96} we evaluate
the correction to the Keldysh Green function
$\delta \hat g^K \propto \bm{v}_F\nabla T$ \cite{FN}.
Combining the resulting expression with Eq. (\ref{current}) we obtain
\begin{gather}
\alpha=-\dfrac{eN_0 v_F^2}{12 T^2}
\int_{-\infty}^{\infty}
\dfrac{\mathcal{F}(\varepsilon)d\varepsilon}{\cosh^2(\varepsilon/2T)}
,
\label{thermo:koeff}
\\
\mathcal{F}(\varepsilon)=\dfrac{ \varepsilon \nu(\varepsilon)\Img\Sigma_0^R(\varepsilon)}{
\left[\Rea\sqrt{\strut\bar \Delta^2 - \bar \varepsilon^2}\right]^2 -
\left[\Img\Sigma_0^R(\varepsilon)\right]^2}.
\label{calF}
\end{gather}

Eqs.  (\ref{thermo:koeff})-(\ref{calF}) -- together with Eqs. (\ref{nusigma0})-(\ref{Gamma2}) -- constitute the central result of this work which
accounts for ``giant'' thermoeffect in superconductors doped by magnetic impurities. In the most relevant case of diffusive
superconductors with $\Gamma \gtrsim T_c$ Eq. (\ref{calF}) reduces to
$\mathcal{F}(\varepsilon)=\nu(\varepsilon)\Img\Sigma_0^R(\varepsilon)/\Gamma^2$, i.e. $\alpha\propto 1/\Gamma^2$ in this limit.

At small magnetic impurity concentrations $\Gamma_2 \ll \Delta$ the impurity band is restricted to subgap
energies $\varepsilon < \Delta$, cf. Fig. \ref{dos-fig}. At energies within the impurity band one has
\begin{equation}
\mathcal{F}(\varepsilon)=
\dfrac{\Gamma_0
\left[2\Gamma_2\Delta\sqrt{1-\beta^2}-(|\varepsilon|-\beta\Delta)^2\right]
}{4\Gamma_2 (1-\beta^2) (\Delta\sqrt{1-\beta^2}+ \Gamma)^2}.
\label{Fsub}
\end{equation}
Substituting this expression into Eq. (\ref{thermo:koeff}), integrating over all impurity band energies and taking the limit
$\Gamma_2 \ll \Delta, T^2/\Delta$, we arrive at the subgap contribution to $\alpha$:
\begin{equation}
\alpha_{sg}=-\dfrac{eN_0 v_F^2}{9 T^2}
\dfrac{\cosh^{-2}(\beta\Delta/2T)\Gamma_0 \sqrt{2\Gamma_2}  \Delta^{3/2}
}{ (1-\beta^2)^{1/4} (\Delta\sqrt{1-\beta^2}+ \Gamma)^2},
\label{alpha_impband}
\end{equation}
i.e. $\alpha_{sg}\propto n_{\rm imp}^{3/2}$ at small concentrations of magnetic impurities.
Assuming that the impurity band is located at $\varepsilon \sim\Delta/2$ ($\beta\sim0.5$) and setting $T \sim \Delta \sim \Gamma_{0,1,2}$
we get
\begin{equation}
\alpha_{sg}=-eN_0 v_F^2T/\Gamma^2 \sim \alpha_N p_F \ell.
\label{alpha_impband2}
\end{equation}
This estimate demonstrates that $\alpha_{sg}$ may strongly exceed the thermoelectric coefficient in the normal state.

Additional contribution to $\alpha$ is provided by overgap energies.
For small values $n_{\rm imp}$ we can use the standard BCS expression for the density of states and derive
\begin{equation}
\mathcal{F}(\varepsilon)=
\dfrac{2\Gamma_0\Gamma_2 (1-\beta^2)\varepsilon^4
\Delta^2 [\varepsilon^2  - \beta^2\Delta^2]^{-1}
}{
\left[
\Gamma_2 \varepsilon^2 +
\Gamma(\varepsilon^2-\beta^2\Delta^2) +
\Gamma_1 (\varepsilon^2-\Delta^2)\right]^2}.
\label{Fover}
\end{equation}
Combining Eqs. (\ref{Fover}) and (\ref{thermo:koeff}),  in a realistic limit $\Gamma\gg\Gamma_{1,2}$  and for $T\sim\Delta$ we recover
the contribution to $\alpha$ from overgap energies
\begin{equation}
\alpha_{og} \sim - e N_0 v_F^2 \dfrac{\Gamma_0\Gamma_2}{\Delta \Gamma^2} \sim \alpha_N\dfrac{\Gamma_0\Gamma_2}{\Delta^2}
p_F\ell.
\end{equation}

In the optimal case $T \sim \Delta \sim \Gamma_{0,1,2,}$ we find $\alpha_{og} \sim \alpha_{sg}$, where the latter quantity obeys
Eq. (\ref{alpha_impband2}). Hence, also for $\alpha=\alpha_{sg}+\alpha_{og}$ we recover the estimate (\ref{est}).

\begin{figure}
\centerline{\includegraphics[width=80mm]{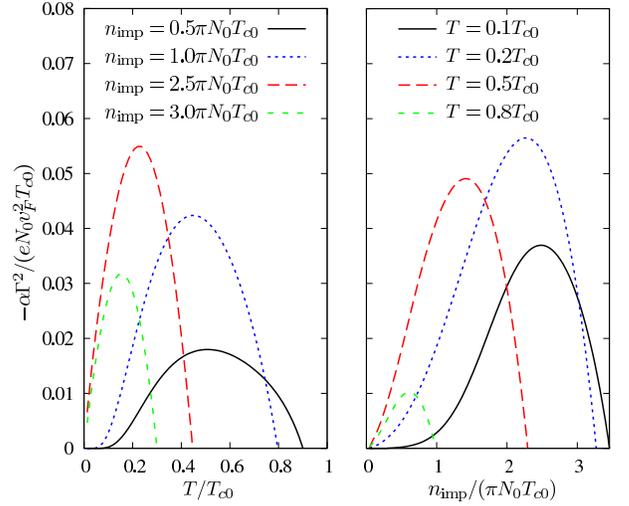}}
\caption{(Color online) Thermoelectric coefficient as a function of temperature and magnetic impurity concentration.
Scattering parameters $u_1=u_2=0.5$ and the scattering rate $\Gamma=10T_{c0}$ are the same for both panels.}
\label{tp-g10-fig}
\end{figure}

At temperatures close to $T_c$ the value $\alpha$ can be evaluated analytically at any concentration of impurities.
In this limit one can set $\nu(\varepsilon)=1$ and obtain
\begin{equation}
\mathcal{F}(\varepsilon)=
\dfrac{2\Gamma_0 \Gamma_2(1-\beta^2)\varepsilon^2\Delta^2 }{(\varepsilon^2+\Gamma_2^2)^2(\Gamma +\Gamma_1 + \Gamma_2)^2},
\end{equation}
which yields
\begin{equation}
\alpha=
-\dfrac{eN_0 v_F^2}{6\pi T^2}
\dfrac{\Gamma_0 (1-\beta^2)\Delta^2 }{(\Gamma +\Gamma_1 + \Gamma_2)^2}
S\left( \dfrac{\Gamma_2}{2\pi T} \right),
\label{therm_Tc}
\end{equation}
where $S(x)=\left[x\psi'\left(x+1/2\right)\right]'$ and  $\psi(x)$ is the digamma function.

The results of numerical evaluation of $\alpha$ as a function of both temperature and impurity concentration are displayed in Fig. \ref{tp-g10-fig}. We observe that the thermoelectric coefficient of a diffusive superconductor
achieves its maximum value at temperatures $T\sim T_c/2$ and $n_{\rm imp}$ approximately equal to one-half of the critical concentration
at which superconductivity gets fully suppressed. This maximum value can be estimated as
\begin{equation}
\max_{T, n_{\rm imp}} |\alpha| \approx 0.05 \dfrac{e N_0 v_F^2 T_{c0}}{\Gamma^2}=
0.2 e N_0 T_{c0} \ell^2 .
\label{alphaSopt}
\end{equation}
Combining the expression for $\alpha_N \sim (\sigma_N/e)(T/\epsilon_F)$ with Eq. \eqref{alphaSopt} we arrive at the estimate (\ref{est})
which demonstrates that enhancement of the thermoeffect is stronger in cleaner superconductors. At the
borderline of applicability of Eq. (\ref{est}) $\ell \sim v_F/T_c$ we obtain
$|\alpha |\sim \sigma_N/e$, which appears to define the absolute maximum value of  $\alpha$ in conventional superconductors doped by magnetic impurities.

It is interesting to point out that the presence of electron-hole asymmetry in such superconductors was also
predicted to yield anomalously large photovoltaic effect \cite{Zaitsev86}. Despite clear similarity
between the models the effect \cite{Zaitsev86} is substantially different from one analyzed here.
Indeed, while no voltage occurs in the system within the linear response to a temperature gradient
\cite{FN}, a non-zero nonequilibrium voltage is induced as a second order response to an external electromagnetic field
\cite{Zaitsev86}. Hence, thermal heating of the system considered here is physically not
equivalent to that produced by an external ac field.

{\it Bimetallic superconducting rings and TEB.} Finally let us briefly discuss the possibility to experimentally
detect ``giant'' thermoeffect predicted here. One way to do so would be to perform an experiment with bimetallic
superconducting rings \cite{Zavaritskii74,Falco76,Harlingen80} as shown in Fig. \ref{ss-fig}. Provided
superconducting contacts are kept at different temperatures $T_a$ and $T_b$, thermoelectric current will
be induced inside the ring and the corresponding magnetic flux $\Phi$ can be measured. The magnitude of
this flux reads
\begin{equation}
\dfrac{\Phi }{\Phi_0}=\dfrac{4e}{c^2}
\int_{T_a}^{T_b}
[\lambda_1^2(T)\alpha_1(T)-\lambda_2^2(T)\alpha_2(T)]dT,
\label{phi-thick}
\end{equation}
where $\Phi_0 = \pi c/e$ is flux quantum, $\alpha_{1,2}$ and $\lambda_{1,2}$ define respectively thermoelectric coefficients and
the values of London penetration depth for two superconductors. For simplicity we may assume $\alpha_{1}\gg \alpha_{2}$ and neglect the second term
in Eq. (\ref{phi-thick}). Employing Eq. \eqref{alphaSopt} together with the standard expression for the London penetration depth
in diffusive superconductors at $T=0$ we arrive at a conservative estimate for the thermally induced flux
\begin{equation}
\dfrac{|\Phi |}{\Phi_0} \sim 0.01 \dfrac{|T_b - T_a|}{\Gamma}, \quad \Gamma \gtrsim T_{c0}.
\end{equation}
\begin{figure}
\centerline{\includegraphics[width=80mm]{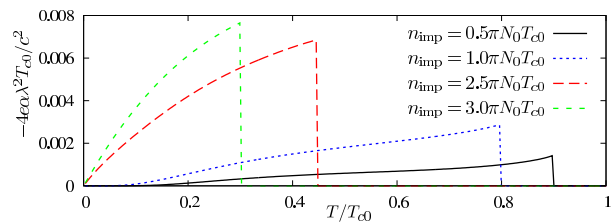}}
\caption{(Color online) Temperature dependence of the term $-4e\alpha \lambda^2T_{c0}/c^2$. Different curves correspond to different
values $n_{\rm imp}$. The parameters $u_{1,2}$ and $\Gamma$ are the same as in Fig. \ref{tp-g10-fig}.}
\label{phi-g10-fig}
\end{figure}

In Fig. \ref{phi-g10-fig} we display the temperature dependence
of the combination $\lambda^2(T)\alpha(T)$ at  different concentrations of magnetic impurities. Induced thermoflux $\Phi$ (normalized to $\Phi_0$)
equals to the area under the corresponding curve between $T_a$ and $T_b$. For reasonably clean superconductors typical values of $\Phi$
may easily reach $\Phi\gtrsim 10^{-2}\Phi_0$.

Another way to experimentally test our predictions would be to employ a novel type of
zero-biased thermo-electric bolometer (TEB) \cite{Kuzmin}.
This TEB consists of a superconducting absorber attached to normal and superconducting electrodes via
tunnel junctions (SIN and SIS' junctions). Incoming photons excite quasiparticles in the absorber.
Strong charge imbalance between excited quasiparticles and quasiholes can be expected provided this absorber
is formed by a superconductor doped by magnetic impurities.
Temperature gradient across the superconductor will occur due to permanent escape of excited quasiparticles from the ``cold''
end of the absorber through SIN junction, whereas no such escape would be possible in its ``hot'' end attached to SIS' junction.
As a result, permanent thermoelectric current will flow in the absorber creating ``giant''
thermoelectric voltage response which can be detected experimentally.

In summary, we have demonstrated that ``giant'' thermoeffect might occur in conventional superconductors doped by
magnetic impurities. This effect is well in the measurable range and can be detected in modern experiments.

The work was supported by the Act 220 of the Russian 
Government (project 25). One of us (A.D.Z.) also 
acknowledges partial support of Deutsche Forschungsgemeinschaft.


\begin{thebibliography}{99}
\bibitem{FN} We emphasize that charge neutrality is explicitly maintained within our approach,
since the induced voltage vanishes identically, $V \propto \langle \Sp \delta \hat g^K\rangle \propto \langle \bm{v}_F\rangle \equiv 0$.
\end{thebibliography}
\end{document}